\documentclass[preprint,showkeys,showpacs,10pt,prd]{revtex4}%

\makeatletter

\renewcommand*{\p@subsection}{}

\renewcommand*{\p@subsubsection}{}
\makeatother

\usepackage[Export]{adjustbox}
\usepackage{amsmath, amssymb, amsfonts, amsthm}
\usepackage{epsfig}
\usepackage{graphicx}  
\usepackage[utf8]{inputenc}
\usepackage{color}     
\usepackage{lscape}
\usepackage{color}
\begin{document}

\title{\Large \bf $TeV$-Scale Resonant Leptogenesis with New Scaling Ansatz on Neutrino Dirac Mass Matrix from $A_4$ Flavor Symmetry}

\author{H. B. Benaoum
} 
\email{hbenaoum@sharjah.ac.ae}
\author{and S. H. Shaglel} 
\affiliation{
Department of Applied Physics and Astronomy, \\
University of Sharjah, United Arab Emirates 
}

\begin{abstract}
We propose a new scaling ansatz in the neutrino Dirac mass matrix to explain the low energy neutrino oscillations data, baryon number asymmetry and neutrinoless double beta decay. In this work, a full reconstruction of the neutrino Dirac mass matrix has been realized from the low energy neutrino oscillations data based on type-I seesaw mechanism. A concrete model based on $A_4$ flavor symmetry has been considered to generate such a neutrino Dirac mass matrix and imposes a relation between the two scaling factors. In this model, the right-handed Heavy Majorana neutrino masses are quasi-degenerate at TeV mass scales. 
Extensive numerical analysis studies have been carried out to constrain the parameter space of the model from the low energy neutrino oscillations data. It has been found that the parameter space of the Dirac mass matrix elements lies near or below the MeV region and the scaling factor $|\kappa_1|$ has to be less than 10. 
Furthermore, we have examined the possibility for simultaneous explanation of both neutrino oscillations data and the observed baryon number asymmetry in the Universe. 
Such an analysis gives further restrictions on the parameter space of the model, thereby explaining the correct neutrino data as well as the baryon number asymmetry via a resonant leptogenesis scenario.
Finally, we show that the allowed space  for the effective Majorana neutrino mass $m_{ee}$ is also constrained in order to account for the observed baryon asymmetry. 
\end{abstract}

.

\maketitle

\section{Introduction}
Non-zero neutrino masses \cite{fukuda}-\cite{abe2} and the observed baryon asymmetry of the Universe constitute evidence of physics beyond the Standard Model (SM) \cite{ade}. Neutrino oscillation experiments indicate that the reactor mixing angle $\theta_{13}$ is small but not zero and the atmospheric neutrino mixing angle $\theta_{23}$ is almost maximum whereas the solar mixing angle $\theta_{12}$ is large but not maximum. Although, only two mass squared differences have been measured, the absolute values of neutrino masses is still unknown and the nature of neutrinos, which could Dirac or Majorana, is still unclear. On the other hand, there are two possible orderings allowed by experiments,i.e. normal hierarchy (NH) with $m_1 < m_2 < m_3$ and inverted hierarchy (IH) with $m_3 < m_1 < m_2$. However, cosmological data from Planck satellite put a tight constrain on the sum of the absolute neutrino masses $\sum\limits_{\alpha=e,\mu,\tau} |m_{\nu_{\alpha}}| < 0.17~eV$ \cite{ade}. As for the phases, the current trend of the data suggest that the Dirac CP phase is constrained for both NH and IH, however, the neutrino oscillation experiments are insensitive to the Majorana CP phases. The present status of neutrino oscillation parameters $\theta_{ij}, \Delta m_{ij}^2$ and $\delta$ can be found in the latest global fit analysis \cite{capozzi,esteban}. \\

On the theoretical side, one of the most well-known economical idea to generate small neutrino masses is through the so-called seesaw mechanism in which extra fermions or scalars are added to the theory at high scale \cite{minkowski}-\cite{valle}. In particular, neutrino masses are generated in type-I seesaw by introducing heavy right-handed Majorana neutrinos to the SM. However, the seesaw mechanism does not explain the neutrino mixing pattern and the neutrino mass hierarchy. One way to overcome these problems of neutrino masses is to augment the SM symmetry with extra discrete flavor symmetry to restrict the form the mass matrices. In this regard, non-abelian favor discrete symmetries play an important role in understanding the physics flavor and have been used extensively in recent years to explain the observed structure of the mixing matrix \cite{king}.
The scale at which this flavor symmetry may manifest itself can, in principle, can be much larger than the typical weak scale, up to GUT scale and even beyond. An interesting  possibility is to have this new flavor symmetry at TeV scale and to be broken at lower energies to residual symmetries in the leptonic sector. 
There are a series of such models based on the symmetry group $S_3, A_4, S_4, A_5$ and other groups with larger orders \cite{altarelli}-\cite{benaoum1}. In particular $A_4$ flavor models are one of the appealing frameworks beyond the SM, in which $A_4$ \cite{babu}-\cite{altarelli2}, being the smallest group with an irreducible triplet representation, enables us naturally to explain the three families of quarks and leptons. \\

In addition to the above unsolved problems, one of the big mysteries in cosmology is the origin of the baryon asymmetry which is the presence of more baryons over anti-baryons in the Universe. Besides the generation of light neutrino masses, the seesaw has another feature which is called leptogenesis to explain the observed asymmetry of the Universe through the decay of heavy Majorana neutrinos \cite{kuzmin}. If this baryon asymmetry comes from leptogenesis, then leptonic CP symmetry must be broken. Therefore, observation of CP violation in the leptonic sector could be a hint towards leptogenesis.  \\

Scaling ansatz in the neutrino sector has been considered by many authors and several phenomenological studies based on this idea have appeared in the literature \cite{scaling}. In this paper, we have introduced a new scaling ansatz in the neutrino Dirac mass matrix. The light neutrino mass matrix is determined through type-I seesaw mechanism. Then, we show how from the knowledge of the light and heavy masses and mixings to reconstruct analytically the full Dirac mass matrix and compute their effective Yukawa couplings. The approach, that we have developed, selects only those Dirac neutrino mass matrices compatible with the low energy physics observables. \\
This new texture of neutrino Dirac mass matrix, with a relation between the two scaling factors, has been realized in a TeV scale seesaw framework with $A_4$ flavor symmetry. We have extensively studied the phenomenological consequences of the model. In particular, the numerical results indicate that the parameter space of the neutrino Dirac mass matrix elements lies near or below the MeV region. Both normal and inverted are analyzed independently in this work. \\

An interesting feature of the model is that that CP violating decays of the heavy right-handed neutrinos produce a lepton asymmetry in the early Universe, which is converted into an observed baryon asymmetry. We have investigated a resonant leptogenesis scenario by having a tiny splitting between the right-handed neutrino masses while keeping the neutrino Dirac mass matrix unchanged. In an attempt to achieve the observed baryon asymmetry, we have used the parameter space of the Dirac neutrino mass matrix $M_D$ and the heavy neutrino mass matrix $M_R$ consistent with the low energy neutrino oscillation data. The resulting parameter space can be tested by already running experiments. \\ 

This paper is organized as follows. In section 2, we introduce the new scaling ansatz in neutrino Dirac mass matrix and show how to reconstruct all the elements of this matrix from the low energy neutrino oscillation data. In section 3, we discuss the model in the framework of type-I seesaw with $A_4$ flavor symmetry that generates such mass matrices $M_D$ and $M_R$. In this framework, section 4 is entirely devoted to the study of the baryon asymmetry of the Universe through resonant leptogenesis. Extensive numerical study, to constrain the parameter space of the model by $3 \sigma$ global fit, has been carried out. Furthermore, detailed numerical analysis on baryon asymmetry from resonant leptogenesis and constraints from neutrinoless beta decay arising from the scaling ansatz on neutrino Dirac mass matrix has been performed and finally summary of the present work is given in section 5. \\ 

\section{Scaling Neutrino Dirac Mass Matrix}
In an attempt to uncover the underlying structure in the neutrino mass matrix, various parameterizations of the neutrino mass matrix have been made in the literature. 
Neutrinos mass matrices with equality between the matrix elements and textures zeros have been proposed to explain the presently available neutrino oscillation data. One particular ansatz, which has been studied, is called scaling. It correlates the matrix elements of the neutrino mass matrix through a scale factor and it can be realized in different ways. \\
In this section, we consider a model independent approach where the ratios of certain elements of the neutrino Dirac mass matrix are equal. The advantage of this choice is to reduce the number of free parameters, thus the neutrino mass matrix gets a simple form and rich predictions. \\

In the basis where the charged lepton mass matrix is diagonal, the most general $3 \times 3$ complex Dirac mass matrix can be parametrized as: 
\begin{eqnarray}
M_D & = & \left( \begin{array}{ccc}
m_{11} & m_{12} & m_{13} \\
m_{21} & m_{22} & m_{23} \\
m_{31} & m_{32} & m_{33} 
\end{array} \right) ~~~~~~~~~~.
\end{eqnarray}
The Dirac neutrino mass matrix has several unknown parameters. To explain the presently available neutrino oscillation and as well as cosmology data, the number of independent parameters has to be reduced. A natural approach is to expect that some of the elements of the mass matrix must be either vanishing or strongly related. \\

In the present work, we consider a new texture where the elements of the Dirac mass matrix obey a scaling invariant pattern. Explicitly, the off-diagonal elements of the Dirac mass matrix are connected to the diagonal elements through the scale factors $\kappa_1$ and $\kappa_2$ as follows: 
\begin{eqnarray}
\frac{m_{12}}{m_{33}} & = & \frac{m_{31}}{m_{22}} = \frac{m_{23}}{m_{11}} = \kappa_1 \nonumber \\
\frac{m_{21}}{m_{33}} & = & \frac{m_{13}}{m_{22}} = \frac{m_{32}}{m_{11}} = \kappa_2
\end{eqnarray}
where $m_{11}, m_{22}, m_{33}, \kappa_1$ and $\kappa_2$ are complex coefficients. \\
Therefore, the Dirac neutrino mass matrix can be written as: 
\begin{eqnarray}
M_D & = & \left( \begin{array}{ccc}
m_{11} & \kappa_1 m_{33} & \kappa_1 m_{22} \\
\kappa_2 m_{33} & m_{22} & \kappa_1 m_{11} \\
\kappa_2 m_{22} & \kappa_2 m_{11} & m_{33} 
\end{array} \right)  ~~~~~~.
\label{scaling}
\end{eqnarray}
Now, instead of taking a diagonal form of the heavy right-handed neutrino mass matrix, we choose it as having a simple form as:
\begin{eqnarray}
M_R & = & f ~\left( \begin{array}{ccc}
1 & 0 & r \\
0 & r & 1 \\
r & 1 & 0 
\end{array} \right)  
\label{mR}
\end{eqnarray}
where the parameters $f$ and $r$ are considered as real numbers. The above right-handed neutrino mass matrix has three eigenvalues. \\
We will show in the next section how this new texture can be realized in $A_4$ flavor symmetry and a type-I seesaw mechanism. \\

Type-I contribution to the light neutrino mass is:
\begin{eqnarray}
M_{\nu} & = & - M_D M_R^{-1} M_D^{T}  ~~~~~~.
\end{eqnarray}
Using equations (\ref{scaling}) and (\ref{mR}),we get: 
\begin{eqnarray}
M_{\nu} & = & - \frac{1}{(1+r^3)} \left(M_{\nu}^{(r=0)} - r M_{\nu}^{(r)} + r^2 M_{\nu}^{(r^2)} \right)
\end{eqnarray}
where, 
\begin{eqnarray}
M_{\nu}^{(r=0)} & = & \frac{1}{f}~\left( \begin{array}{ccc}
m_{11}^2 + 2 \kappa_1 \kappa_2~ m_{22} m_{33} & \kappa_2~ m_{22}^2 + (\kappa_1^2 + \kappa_2)~ m_{11} m_{33} & \kappa_1~ m_{33}^2 + (\kappa_1 + \kappa_2^2)~ m_{11} m_{22} \\
\kappa_2~ m_{22}^2 + (\kappa_1^2 + \kappa_2)~ m_{11} m_{33} & \kappa_2^2~ m_{33}^2 + 2 \kappa_1~ m_{11} m_{22} & \kappa_1 \kappa_2~ m_{11}^2 + (1+ \kappa_1 \kappa_2)~ m_{22} m_{33} \\
\kappa_1~ m_{33}^2 + (\kappa_1 + \kappa_2^2)~ m_{11} m_{22} & \kappa_1 \kappa_2~ m_{11}^2 + (1+ \kappa_1 \kappa_2)~ m_{22} m_{33} & \kappa_1^2~ m_{22}^2 + 2 \kappa_2 m_{11} m_{33}
\end{array} \right)  ~~~~~~~,
\end{eqnarray}
\begin{eqnarray}
M_{\nu}^{(r)} & = & \frac{1}{f}~\left( \begin{array}{ccc}
\kappa_2^2~ m_{22}^2 + 2 \kappa_1 \kappa_2~ m_{11} m_{33} & \kappa_1 \kappa_2~ m_{33}^2 + (1+ \kappa_1 \kappa_2)~ m_{11} m_{22} & \kappa_2^2~ m_{11}^2 + (\kappa_1^2 + \kappa_2)~ m_{22} m_{33} \\
\kappa_1 \kappa_2~ m_{33}^2 + (1+ \kappa_1 \kappa_2)~ m_{11} m_{22} & \kappa_1^2~ m_{11}^2 + 2 \kappa_2~ m_{22} m_{33} & \kappa_1~ m_{22}^2 + (\kappa_1 + \kappa_2^2)~ m_{11} m_{33} \\
\kappa_2^2~ m_{11}^2 + (\kappa_1^2 + \kappa_2)~ m_{22} m_{33} & \kappa_1~ m_{22}^2 + (\kappa_1 + \kappa_2^2)~ m_{11} m_{33} & m_{33}^2 + 2 \kappa_1 \kappa_2~ m_{11} m_{22}
\end{array} \right)  ~~~~~~~,
\end{eqnarray}
\begin{eqnarray}
M_{\nu}^{(r^2)} & = & \frac{1}{f}~\left( \begin{array}{ccc}
\kappa_1^2~ m_{33}^2 + 2 \kappa_2~ m_{11} m_{22} & \kappa_1~ m_{11}^2 + (\kappa_1 + \kappa_2^2)~ m_{22} m_{33} & \kappa_1 \kappa_2~ m_{22}^2 + (1+ \kappa_1 \kappa_2)~ m_{11} m_{33} \\
\kappa_1~ m_{11}^2 + (\kappa_1 + \kappa_2^2)~ m_{22} m_{33} & m_{22}^2 + 2 \kappa_1 \kappa_2~ m_{11} m_{33} & \kappa_2~ m_{33}^2 + (\kappa_1^2 + \kappa_2)~ m_{11} m_{22} \\
\kappa_1 \kappa_2~ m_{22}^2 + (1+ \kappa_1 \kappa_2)~ m_{11} m_{33} & \kappa_2~ m_{33}^2 + (\kappa_1^2 + \kappa_2)~ m_{11} m_{22} & \kappa_2^2~ m_{11}^2 + 2 \kappa_1 \kappa_2~ m_{22} m_{33}
\end{array} \right)  ~~.
\end{eqnarray}

In the charged lepton basis, the light neutrino mass matrix can be diagonalized by:
\begin{eqnarray}
M_{\nu} & = &  U_{PMNS} M_{\nu}^{diag} U_{PMNS}^T
\end{eqnarray}
where $M_{\nu}^{diag}=diag \left(m_1, m_2, m_3 \right)$ is the diagonal matrix containing the three neutrino masses $m_{1,2,3}$ and $U_{PMNS}= U P_{Maj}$ is the Pontecorvo-Maki-Nakagawa-Sakata unitary matrix given by: 
\begin{eqnarray}
U & = & \left( \begin{array}{ccc}
c_{12} c_{13} & s_{12} c_{13} & s_{13} e^{-i \delta} \\
- s_{12} c_{23} - c_{12} s_{23} s_{13} e^{i \delta} & c_{12} c_{23} - s_{12} s_{23} s_{13} e^{i \delta} & s_{23} c_{13} \\
s_{12} s_{23} - c_{12} c_{23} s_{13} e^{i \delta} & -c_{12} s_{23} - s_{12} c_{23} s_{13} e^{i \delta} & c_{23} c_{13} 
\end{array} \right) 
\end{eqnarray}
where $c_{ij} = \cos \theta_{ij}, s_{ij} = \sin \theta_{ij}, \theta_{ij}$'s are flavor mixing angles; $i,j=1,2,3$; $\delta$ is the Dirac CP-violating phase and $P_{Maj} = diag \left(1, e^{i \alpha/2}, e^{i \beta/2} \right)$ is the diagonal matrix that contains the Majorana phases $\alpha$ and $\beta$. \\

Using the type-I seesaw formula, it is convenient to write the symmetric neutrino mass matrix as: 
\begin{eqnarray}
M_{\nu} & = & M_{\nu}^{(1)} + M_{\nu}^{(2)}  ~~~~~,
\end{eqnarray}
where we have decomposed the neutrino mass matrix into two pieces where $M_{\nu}^{(1)}$ is a symmetric matrix given by:
\begin{eqnarray}
M_{\nu}^{(1)} & = & \left( \begin{array}{ccc}
M_{\nu_{11}} & M_{\nu_{33}} & M_{\nu_{22}} \\
M_{\nu_{33}} & M_{\nu_{22}} & M_{\nu_{11}} \\
M_{\nu_{22}} & M_{\nu_{11}} & M_{\nu_{33}} 
\end{array} \right)
\end{eqnarray}
and $M_{\nu}^{(2)}$ mass matrix with zero diagonal elements having the form,
\begin{eqnarray}
M_{\nu}^{(2)} & = & \left( \begin{array}{ccc}
0 & M_{\nu_{12}} -M_{\nu_{33}} & M_{\nu_{13}} -M_{\nu_{22}}  \\
M_{\nu_{12}} - M_{\nu_{33}}& 0 & M_{\nu_{23}} - M_{\nu_{11}} \\
M_{\nu_{13}} - M_{\nu_{22}}& M_{\nu_{23}} - M_{\nu_{11}} & 0
\end{array} \right)  ~~~~~~.
\end{eqnarray}
Explicit structure of the off-diagonal elements of the mass matrix $M_{\nu}^{(2)}$ can be expressed as follows: 
\begin{eqnarray}
\left( \begin{array}{c} 
M_{\nu_{13}}^{(2)} \\
M_{\nu_{12}}^{(2)} \\
M_{\nu_{23}}^{(2)}
\end{array} \right) & = & \frac{1}{f (1+r^3)} \left( \begin{array}{ccc}
\left(1 - \kappa_1 \kappa_2 \right) ~r^2 &  \kappa_2^2 - \kappa_1 & - \left(\kappa_1^2 - \kappa_2 \right) ~r  \\ 
\kappa_1^2 - \kappa_2 & - \left(1 - \kappa_1 \kappa_2 \right) ~r  & \left(\kappa_2^2 - \kappa_1 \right) ~r^2  \\
-\left(\kappa_2^2 - \kappa_1 \right) ~r & \left( \kappa_1^2 - \kappa_2 \right) ~r^2 & 1 - \kappa_1 \kappa_2 \end{array} \right) 
\left( \begin{array}{c} 
\gamma_3 \\
\gamma_2 \\
\gamma_1 \end{array} \right) = 
\left( \begin{array}{c} 
w_{13} \\
w_{12} \\
w_{23} \end{array} \right) 
\end{eqnarray}
where we have defined $\gamma_1, \gamma_2$ and $\gamma_3$ as,
\begin{eqnarray}
\gamma_1 & = & m_{11}^2 - m_{22} ~m_{33} \nonumber \\
\gamma_2 & = & m_{22}^2 - m_{11} ~m_{33} \nonumber \\
\gamma_3 & = & m_{33}^2 - m_{11} ~m_{22} 
\end{eqnarray} 
which is a convenient way to parametrize the matrix elements of the Dirac mass matrix. Moreover,$w_{12}, w_{13}$ and $w_{23}$ can be expressed in terms of the mixing angles, neutrino masses and CP phases as: 
\begin{eqnarray}
w_{12} & = & \sum_{i=1}^{3} \mu_i \left( U_{1i} U_{2i} - U_{3i}^2 \right) \nonumber \\
w_{13} & = & \sum_{i=1}^{3} \mu_i \left( U_{1i} U_{3i} - U_{2i}^2 \right) \nonumber \\ 
w_{23} & = & \sum_{i=1}^{3} \mu_i \left( U_{2i} U_{3i} - U_{1i}^2 \right) 
\end{eqnarray}
which are determined from the neutrino oscillation data (up to the CP-violating Majorana phases). Here, we have defined the masses including the corresponding Majorana phases as $\mu_1 = m_1, \mu_2 = m_2 e^{i \alpha}$ and $\mu_3 = m_3 e^{i \beta}$. \\

It turns out that the parameters $\gamma_1, \gamma_2$ and $\gamma_3$ can be reconstructed from the six neutrino oscillations parameters, the lightest neutrino mass, two Majorana phases, nonzero entry in the right-handed neutrino mass and scaling factors as:
\begin{eqnarray}
\frac{\gamma_1}{f (1+r^3)} & = & \frac{\left(1+r^3 \right) \left( (1- \kappa_1 \kappa_2 )^2 w_{23} + r (\kappa_2^2 - \kappa_1)^2 w_{12} \right) 
- \left(w_{23} + \kappa_2 r^4 w_{12} - \kappa_1 r^2 w_{13} \right) \kappa_{\omega}}
{(1+r^3)^2 (\kappa_1^2 - \kappa_2) (1- \kappa_1 \kappa_2) (\kappa_2^2 - \kappa_1) + \kappa_{\omega}^2 r^3} \nonumber \\
\frac{\gamma_2}{f (1+r^3)} & = & \frac{\left(1+r^3 \right) \left( (\kappa_1^2- \kappa_2 )^2 w_{12} - r (1- \kappa_1 \kappa_2)^2 w_{13} \right) 
- \left(\kappa_1 w_{12} + r^4 w_{13} + \kappa_2 r^2 w_{23} \right) \kappa_{\omega}}
{(1+r^3)^2 (\kappa_1^2 - \kappa_2) (1- \kappa_1 \kappa_2) (\kappa_2^2 - \kappa_1) + \kappa_{\omega}^2 r^3} \nonumber \\
\frac{\gamma_3}{f (1+r^3)} & = & \frac{\left(1+r^3 \right) \left( (\kappa_2^2- \kappa_1)^2 w_{13} + r (\kappa_1^2 - \kappa_2)^2 w_{23} \right) 
- \left(\kappa_2 w_{13} - \kappa_1 r^4 w_{23} + r^2 w_{12} \right) \kappa_{\omega}}
{(1+r^3)^2 (\kappa_1^2 - \kappa_2) (1- \kappa_1 \kappa_2) (\kappa_2^2 - \kappa_1) + \kappa_{\omega}^2 r^3} 
\label{gamma}
\end{eqnarray}
where we have defined $\kappa_{\omega}$ as: 
\begin{eqnarray}
\kappa_{\omega} = (\kappa_1 + \kappa_2 + 1) (\kappa_1 +  \omega + \omega^2 \kappa_2 ) (\kappa_1 + \omega^2 + \omega \kappa_2 )  ~~~~~.
\end{eqnarray}

It is interesting to note that the full Dirac mass matrix, which is described altogether by 10 real parameters, can be reconstructed analytically, in the most general case, as: 
\begin{eqnarray}
m_{11} & = & \pm \frac{\gamma_1^2 - \gamma_2 \gamma_3}{\sqrt{\gamma_1^3 - 3 \gamma_1 \gamma_2 \gamma_3 + \gamma_2^3 + \gamma_3^3}} \nonumber \\
m_{22} & = & \pm \frac{\gamma_2^2 - \gamma_1 \gamma_3}{\sqrt{\gamma_1^3 - 3 \gamma_1 \gamma_2 \gamma_3 + \gamma_2^3 + \gamma_3^3}} \nonumber \\
m_{33} & = & \pm \frac{\gamma_3^2 - \gamma_1 \gamma_2}{\sqrt{\gamma_1^3 - 3 \gamma_1 \gamma_2 \gamma_3 + \gamma_2^3 + \gamma_3^3}}  ~~~~~.
\label{elements}
\end{eqnarray}

We can get further two more relations between the scaling factors from the determinant and the trace of the neutrino mass matrix $M_{\nu}$ as: 
\begin{eqnarray}
Det \left(M_{\nu} \right) & = & \frac{1}{f^3 \left(1+r^3\right)} ~\frac{\left(\kappa_{\omega} (\gamma_1^2 - \gamma_2 \gamma_3) (\gamma_2^2 - \gamma_1 \gamma_3) 
(\gamma_3^2 - \gamma_1 \gamma_2) - \kappa_1 \kappa_2 (\gamma_1^3 + \gamma_2^3 + \gamma_3^3 -3 \gamma_1 \gamma_2 \gamma_3)^2 \right)^2}{(\gamma_1^3 + \gamma_2^3 
+\gamma_3^3 -3 \gamma_1 \gamma_2 \gamma_3)^3} \nonumber \\
= \mu_1 \mu_2 \mu_3 \nonumber \\
Tr \left(M_{\nu} \right) & = & \frac{N_{\nu}}{f \left(1+r^3\right) (\gamma_1^3 + \gamma_2^3 + \gamma_3^3 -3 \gamma_1 \gamma_2 \gamma_3)} 
= \mu_1 + \mu_2 + \mu_3 
\label{tracedet}
\end{eqnarray}
where 
\begin{eqnarray}
N_{\nu} & = & - \left(\gamma_1^2 - \gamma_2 \gamma_3 + \kappa_1 (\gamma_2^2 - \gamma_1 \gamma_3) + \kappa_2 (\gamma_3^2 - \gamma_1 \gamma_2) \right)^2 + 
r \left(\gamma_3^2 - \gamma_1 \gamma_2 + \kappa_1 (\gamma_1^2 - \gamma_2 \gamma_3) + \kappa_2 (\gamma_2^2 - \gamma_1 \gamma_3) \right)^2  \nonumber \\
& & - r^2 \left(\gamma_2^2 - \gamma_1 \gamma_3 + \kappa_1 (\gamma_3^2 - \gamma_1 \gamma_2) + \kappa_2 (\gamma_1^2 - \gamma_2 \gamma_3) \right)^2  ~~~.
\end{eqnarray}

From equations (\ref{gamma}), (\ref{elements}) and (\ref{tracedet}), all the elements of $M_D$ are determined from neutrino oscillation data, lightest neutrino mass, $f$ and $r$. With this method, we compute all Dirac neutrino mass matrices that satisfy the oscillation constraints, and conduct further study of the TeV-scale phenomenology of these matrices, focusing on leptogenesis. \\

In the following section, we present an $A_4$ discrete symmetry model with scaling invariant texture for the neutrino Dirac mass $M_D$ with 
$\kappa_1 + \kappa_2=-1$ that leads to a resonant leptogenesis with TeV-scale type-I seesaw in a natural way, thereby explaining the correct neutrino data as well as the baryon asymmetry.
\section{$A_4$ Realization of the Model}
Here, we consider the Standard Model group with an additional global flavor symmetry $A_4$. 
The group $A_4$ is a group that describes even permutations of four objects. It has 12 elements, which can be generated by two basic generators $S$ and $T$ satisfying the relations, 
\begin{eqnarray*}
S^2 & = & T^3 = \left(S T \right)^3 = 1   ~~.
\end{eqnarray*}   
The group representations of $A_4$ are relatively simple and it has four in-equivalent irreducible representations, including three one-dimensional representations $1, 1', 1''$,
\begin{eqnarray*}
1:~~~~~S=1,~~~~~~T =1 \nonumber\\
1':~~~~~S=1,~~~~~~T = \omega \nonumber\\
1'':~~~~~S=1,~~~~~~T = \omega^2 \nonumber\\
\end{eqnarray*}
and one three-dimensional representation generated via $S$ and $T$, 
\begin{eqnarray*}
T  =  \left( \begin{array}{ccc}
1 & 0 & 0 \\
0 & \omega^2 & 0 \\
0 & 0 & \omega 
\end{array} \right)~~,  ~~~~~~~S  =  \frac{1}{3}~ \left( \begin{array}{ccc}
-1 & 2 & 2 \\
2 & -1 & 2 \\
2 & 2 & -1
\end{array} \right) ~~~~~~~.
\end{eqnarray*}
The product rules of the irreducible representations are given as: 
\begin{eqnarray*}
1 \otimes 1 & = & 1 \nonumber \\
1 \otimes 1' & = & 1'' \nonumber \\
1' \otimes 1'' & = & 1 \nonumber \\
1'' \otimes 1'' &  = & 1' \nonumber \\
3 \otimes 3 & = & 1 \oplus 1' \oplus 1'' \oplus 3_a \oplus 3_s  ~~~~~.
\end{eqnarray*} 
This direct product can be decomposed into three singlets and two triplets as follows: 
\begin{eqnarray*}
\left(3 \otimes 3 \right)_1 & = & a_1 b_1 + a_2 b_2 + a_3 b_3 \\
\left(3 \otimes 3 \right)_{1'} & = & a_1 b_2 + a_2 b_1 + a_3 b_3 \\
\left(3 \otimes 3 \right)_{1''} & = & a_1 b_3 + a_2 b_2 + a_3 b_1 \\
\left(3 \otimes 3 \right)_{3_a} & = & \frac{1}{2}~ \left(a_2 b_3 - a_3 b_2, a_1 b_2 - a_2 b_1, a_3 b_1 - a_1 b_3  \right) \\
\left(3 \otimes 3 \right)_{3_s} & = & \frac{1}{3}~ \left(2 a_1 b_1 - a_2 b_3 - a_3 b_2, 2 a_3 b_3 - a_1 b_2 - a_2 b_1, 2 a_2 b_2 - a_1 b_3 - a_3 b_1 \right)
\end{eqnarray*}
where $\omega = e^{2 \pi i/3}$ and $(a_1,a_2,a_3), (b_1,b_2,b_3)$ are basis vectors of the two triplets. \\

We consider a discrete symmetry based on a natural type-I seesaw in this section. The Dirac neutrino mass matrix with scaling ansatz, although it looks {\it ad hoc}, can however be realized in this framework using generic flavor symmetry models like $A_4$. \\
We will use the above prescription to construct the Lagrangian. In addition to the three standard lepton $SU(2)_L$ doublets $L_{\alpha}$ (where $\alpha$ stands for generations), three right-handed charged lepton singlet $e_R, \mu_R$ and $\tau_R$, and $SU(2)_L$ doublet Higgs scalar $H$, we extend the SM by three copies of right-handed singlet neutrinos $N_i; i \in 1,2,3$, and four $SU(2)_L$ scalar singlet fields $\phi_E, \phi_{\nu}, \xi$ and $\zeta$. In this model, the right-handed neutrinos $N_i$, the scalars $\phi_E$ and $\phi_{\nu}$ form a triplet under $A_4$ group whereas $\xi$ and $\xi'$ transform as an $A_4$ singlets $1$ and $1'$ respectively. We also impose $\mathbb{Z}_3 \times \mathbb{Z}_2$ as an additional symmetry to prevent coupling between charged leptons and $SU(2)_L$ singlet scalars. 
The transformation properties of various fields under $A_4$ are given in Table  \ref{tab:Table 1}.\\

\begin{table}[h]
\begin{center}
\renewcommand{\arraystretch}{1.5}
\begin{tabular}{ccccccccccc}
\hline
\hline
 & $\bar{L}$ & $e_R$ & $\mu_R$ & $\tau_R$ & $N$ & $H$ & $\phi_E$ & $\phi_{\nu}$ & $\xi$ & $\xi'$ \\
\hline
$SU(2)_L$  & $2$ & $1$ & $1$ & $1$ & $1$ & $2$ & $1$ & $1$ & $1$ & $1$ \\
$A_4$     & $3$ & $1$ & $1'$ & $1''$ & $3$ & $1$ & $3$ & $3$ & $1$ & $1''$ \\
$Z_3$     & $\omega$ & $\omega^2$ & $\omega^2$ & $\omega^2$ & $\omega$ & $1$ & $1$ & $\omega$ & $\omega$ & $1$ \\
$Z_2$     & $1$ & $1$ & $1$ & $1$ & $-1$ & $1$ & $1$ & $-1$ & $1$ & $-1$ \\
\hline
\hline
\end{tabular}
\caption{\label{tab:Table 1} Fields assignments under $SU(2)_L$ and $A_4$ symmetry.}
\end{center}
\end{table}

Based on the $A_4 \times \mathbb{Z}_3 \times \mathbb{Z}_2$ symmetry, we construct the following Yukawa terms of the effective Lagrangian for the lepton sector,
\begin{eqnarray}
{\cal L} & = & y_e \bar{L} H \frac{\phi_E}{\Lambda} e_R + y_{\mu} \bar{L} H \frac{\phi_E}{\Lambda} \mu_R + y_{\tau} \bar{L} H \frac{\phi_E}{\Lambda} \tau_R 
+ \frac{y_s}{\Lambda} \left(\phi_{\nu} \bar{L} \right)_{3_s} \tilde{H} N + \frac{y_a}{\Lambda} \left(\phi_{\nu} \bar{L} \right)_{3_a} \tilde{H} N \nonumber \\
& + & y_N \left(N N \right)_1 \xi + y'_N \left(N N \right)_{1''} \xi \frac{\xi' \xi'}{\Lambda^2} + H.c.
\end{eqnarray}
where $\tilde{H}$ is the conjugate of $H$ related by $\tilde{H} = i \tau_2 H^*$.  \\
The Lagrangian for the neutrino sector has two contributions from the triplets in the symmetry and anti-symmetric product $3 \otimes 3$ and two others from the singlets $1$ and $1'$. The anti-symmetric part of the Dirac mass term is crucial for generating non-zero mixing angle $\theta_{13}$ \cite{memenga}. \\

By denoting the vacuum expectation values (vevs) of the Higgs boson $H$ to be $<H>=v_H$, the singlet scalars $<\xi>=u$ and $<\xi'>=u'$ and choosing specific vev alignments of the triplet scalars as:
\begin{eqnarray}
<\phi_E>  =  \left(v_E, 0, 0 \right), ~~~<\phi_{\nu}>  =  \left(v_{\nu_1}, v_{\nu_1}, v_{\nu_1} \right) ~~~.
\end{eqnarray}
The resulting neutrino mass matrix has three eigenvalues, where one of them is zero and the two others are degenerate, which is in contradiction to the neutrino oscillation experiments. It also gives a non-vanishing reactor angle $\theta_{13}$ that deviates from tri-bimaximal mixing such that $\sin^2 \theta_{12}=1/3, \sin^2 \theta_{23} =1/2$ and $\sin^2 \theta_{13} \neq 0$. \\

In order to have the correct low energy phenomenology, we choose a more general vacuum alignment for the scalar triplet $<\phi_{\nu}>  =  \left(v_{\nu_1}, v_{\nu_2}, v_{\nu_3} \right)$ which can be considered as a small perturbation around the exact vev alignment. Such a choice corrects the neutrino mass spectrum and gives rise to the correct mass squared differences as well as mixing angles. The above vev configuration is a solution of the minimization conditions of the scalar potential provided provided it softly breaks the flavor discrete symmetry. Such a deviation is being then associated to this soft breaking. The way how the above vacuum configuration is achieved is out of the scope of the present work.  \\

The resulting form of the charged lepton mass matrix arising from the model, after symmetry breaking, is diagonal given by:
\begin{eqnarray}
M_l & = & \frac{v_H v_E}{\Lambda} \left( \begin{array}{ccc}
y_e & 0 & 0 \\
0 & y_{\mu} & 0 \\
0 & 0 & y_{\tau} 
\end{array} \right)  ~~~~.
\end{eqnarray}
The chosen vevs allow us to have a scaling invariant Dirac neutrino mass matrix,
\begin{eqnarray}
M_D & = & \frac{v_H}{\Lambda} \left( \begin{array}{ccc}
\frac{2}{3} y_s v_{\nu 1} & - \left(\frac{y_s}{3} + \frac{y_a}{2} \right) v_{\nu 3} & - \left(\frac{y_s}{3} - \frac{y_a}{2} \right) v_{\nu 2} \\
- \left(\frac{y_s}{3} - \frac{y_a}{2} \right) v_{\nu 3} &\frac{2}{3} y_s v_{\nu 2}  & - \left(\frac{y_s}{3} + \frac{y_a}{2} \right) v_{\nu 1} \\
- \left(\frac{y_s}{3} + \frac{y_a}{2} \right) v_{\nu 2} & - \left(\frac{y_s}{3} - \frac{y_a}{2} \right) v_{\nu 1} & \frac{2}{3} y_s v_{\nu 3} 
\end{array} \right)
\end{eqnarray}
and a right-handed heavy neutrino mass matrix which can be written as:
\begin{eqnarray}
M_R & = & u ~\left( \begin{array}{ccc}
2 y_N & 0 & y'_N \frac{{u'}^2}{\Lambda^2} \\
0 & y'_N \frac{{u'}^2}{\Lambda^2} & 2 y_N \\
y'_N \frac{{u'}^2}{\Lambda^2} & 2 y_N & 0 
\end{array}
\right)  ~~~~.
\end{eqnarray}
Hence, the above two matrices take the following form:
\begin{eqnarray}
M_D & = & \left( \begin{array}{ccc}
a & \kappa_1 d & \kappa_2 c \\
\kappa_2 d & c & \kappa_1 a \\
\kappa_1 c & \kappa_2 a & d 
\end{array} \right)
\end{eqnarray}
and 
\begin{eqnarray}
M_R & = & f~ \left( \begin{array}{ccc}
1 & 0 & r \\
0 & r & 1 \\
r & 1 & 0 
\end{array}
\right)
\end{eqnarray}
where $a=\frac{2}{3} y_s \frac{v_{\nu 1} v_H}{\Lambda}, c=\frac{2}{3} y_s \frac{v_{\nu 2} v_H}{\Lambda}, d=\frac{2}{3} y_s \frac{v_{\nu 3} v_H}{\Lambda}, 
f = 2 y_N u, r = \frac{{u'}^2}{2 \Lambda^2} \frac{y'_N}{y_N}, \kappa_1 = -\frac{1}{2} - \frac{3 y_a}{4 y_s}$ and 
$\kappa_2 = -\frac{1}{2} + \frac{3 y_a}{4 y_s}$; with $\kappa_1 + \kappa_2 = -1$. \\

Now, the right handed neutrino mass matrix $M_R$ has two parameters $f$ and $r$ where $f$ is a leading term and $r$ is a parameter creating a tiny mass splitting needed to give rise to a successful leptogenesis. The contribution to the Lagrangian $(23)$ from dimensions 5 allowed by the symmetry that spoils the structure of the right handed neutrino $M_R$ mass matrix can arise, for example,by adding dimension 5 like $NN \phi_{\nu} \xi'/\Lambda$ which gives a contribution $\delta M_R$ to $M_R$ that has a leading order $f$,
\begin{eqnarray*}
M_R + \delta M_R & = & M_R ~\left(1+ M_R^{-1} \delta M_R \right) = M_R ~\left(1 + O(\frac{v_{\nu 1} u'}{f \Lambda}) \right) \simeq M_R  ~~.
\end{eqnarray*}
Such corrections can lead, in principle, to deviations from the trivial form of the right handed neutrino mass matrix but the relevance of such term in the actual numerical calculations is negligible compared to the leading order $f$. \\

The above $A_4$ flavor symmetry model gives a relation between the two scaling factors, i.e. $\kappa_1 + \kappa_2= -1$. From the equation (19) and imposing the condition $\kappa_{\omega}=0$, we see that besides the condition $\kappa_1 + \kappa_2=-1$, there are two others relations between the scaling factors, namely, 
$\omega^2 \kappa_1 + \omega \kappa_2 = -1$ and $\omega \kappa_1 + \omega^2 \kappa_2=-1$. These two relations correspond to different models than the one considered here and may lead to interesting theoretical and phenomenological implications. \\

The neutrino mass matrix $M_{\nu}^{(2)}$ with zero-diagonal entries becomes, 
\begin{eqnarray}
M_{\nu}^{(2)} & = & \frac{1}{f (1+r^3) (\kappa_1^2 + \kappa_1 +1)}~\left( \begin{array}{ccc}
0 & \gamma_2 - \gamma_3 r + \gamma_1 r^2 & \gamma_3 - \gamma_1 r + \gamma_2 r^2  \\
\gamma_2 - \gamma_3 r + \gamma_1 r^2 & 0 & \gamma_1 - \gamma_2 r + \gamma_3 r^2\\
\gamma_3 - \gamma_1 r + \gamma_2 r^2 & \gamma_1 - \gamma_2 r + \gamma_3 r^2 & 0
\end{array} \right)   ~~~~~~.
\end{eqnarray}
Since $\kappa_2=-1-\kappa_1$, the parameters $\gamma_1, \gamma_2$ and $\gamma_3$ take a simple form:
\begin{eqnarray}
\gamma_1 & = & f~ \frac{(1- \kappa_1 \kappa_2 )^2 w_{23} + r (\kappa_2^2 - \kappa_1)^2 w_{12}} 
{(\kappa_1^2 - \kappa_2) (1- \kappa_1 \kappa_2) (\kappa_2^2 - \kappa_1)} = f~ \frac{w_{23} + r w_{12}} 
{\kappa_1^2 + \kappa_1 +1} \nonumber \\
\gamma_2 & = & f~\frac{(\kappa_1^2- \kappa_2 )^2 w_{12} - r (1- \kappa_1 \kappa_2)^2 w_{13}}
{(\kappa_1^2 - \kappa_2) (1- \kappa_1 \kappa_2) (\kappa_2^2 - \kappa_1)} = f~\frac{w_{12} - r w_{13}}
{\kappa_1^2 + \kappa_1 +1} \nonumber \\
\gamma_3 & = & f~\frac{(\kappa_2^2- \kappa_1)^2 w_{13} + r (\kappa_1^2 - \kappa_2)^2 w_{23}}
{(\kappa_1^2 - \kappa_2) (1- \kappa_1 \kappa_2) (\kappa_2^2 - \kappa_1)} = f~\frac{w_{13} + r w_{23}}
{\kappa_1^2 + \kappa_1 + 1} ~~~~~~~.
\end{eqnarray}
The determinant of the light active neutrino mass matrix $M_{\nu}$ can now be written as: 
\begin{eqnarray}
Det \left(M_{\nu} \right) & = & \frac{\kappa_1^2 (1+\kappa_1)^2}{f^3 (1+r^3)} \left(\gamma_1^3 + \gamma_2^3 + \gamma_3^3 -3 \gamma_1 \gamma_2 \gamma_3 \right) 
= \mu_1 \mu_2 \mu_3 
\end{eqnarray} 
Then one obtains the equation that constrains the scaling parameter in terms of our inputs, 
\begin{eqnarray}
\frac{\left(\kappa_1^2 + \kappa_1 +1 \right)^3}{\kappa_1^2 \left(1+\kappa_1 \right)^2} & = & \frac{(w_{23} +r w_{12})^3+(w_{12}-r w_{13})^3+(w_{13}+r w_{23})^3- 3 (w_{23}+r w_{12}) (w_{12}- r w_{13}) (w_{13}+ r w_{23})}{(1+r^3) \mu_1 \mu_2 \mu_3}
\end{eqnarray}
The solution of the above expression determines the scaling factor $\kappa_1$ as a function of $m_{\small{lightest}}, \Delta m_{\odot}^2 , \Delta m_{atm}^2 $,
$\sin^2 \theta_{12}, \sin^2 \theta_{13}, \sin^2 \theta_{23}, \delta, \alpha$, $\beta$ and $M_R$. Detailed analysis will be performed in section \ref{section5}.
\section{Leptogenesis}
Leptogenesis, which is a cosmological consequence of the minimal type-I seesaw mechanism, is an elegant framework to explain the observed baryon asymmetry of the universe. A lepton asymmetry is generated from the CP-violating out-of-equilibrium decays of the heavy Majorana right-handed neutrinos. To realize a thermal leptogenesis at TeV scales in the Universe, a mass degeneracy among the heavy right-handed neutrinos is required. \\

Here, we will investigate the the baryogenesis via resonant leptogenesis where the TeV heavy right-handed neutrinos have nearly degenerate masses and the CP-asymmetries are due to the self-energy effects in the heavy right-handed Majorana neutrino decays. \\

The heavy Majorana neutrino mass matrix $M_R$ has the small parameter $r$ which is a crucial for the resonant leptogenesis~\cite{pilaftsis}. The right-handed neutrino mass matrix has three eigenvalues where the parameter $f$ is the leading order right-handed mass and $r$ is the parameter creating tiny mass splitting. The full mass matrix $M_R$ can be diagonalized as: 
\begin{eqnarray}
M_R & = & U_R M_R^{diag} U_R^T 
\end{eqnarray}
where the mass eigenvalues are given by:
\begin{eqnarray}
M_R^{diag} & = & diag \left(M_1 = f \sqrt{1-r+r^2}, M_2 = f (1+r), M_3 = - f \sqrt{1-r+r^2} \right)  ~~~.
\end{eqnarray}
The unitary matrix $U_R$ can be expressed as: 
\begin{eqnarray}
U_R & = & U_{TBM} U_{R~13} 
\end{eqnarray}
where
\begin{eqnarray}
U_{TBM} & = & \left( \begin{array}{ccc}
\sqrt{\frac{2}{3}} & \frac{1}{\sqrt{3}} & 0 \\
- \frac{1}{\sqrt{6}}& \frac{1}{\sqrt{3}} & -\frac{1}{\sqrt{2}} \\
- \frac{1}{\sqrt{6}} & \frac{1}{\sqrt{3}} &  \frac{1}{\sqrt{2}}
\end{array}
\right)
\end{eqnarray}
and 
\begin{eqnarray}
U_{R~13} & = & \left( \begin{array}{ccc}
\cos \theta_R & 0 & \sin \theta_R \\
0 & 1 & 0 \\
- \sin \theta_R & 0 & \cos \theta_R 
\end{array}
\right)
\end{eqnarray}
with $\tan 2 \theta_R = \frac{\sqrt{3} r}{2-r}$. When $r=0$, the unitary matrix $U_R$ is the pure tri-bimaximal matrix and non-zero for the parameter $r$ encodes deviation from tri-bimaximal pattern. \\

Consequently, the Yukawa $Y_{\nu}$ of the Dirac mass matrix in the diagonal basis of $M_R$ is: 
\begin{eqnarray}
Y_{\nu} & = & \frac{1}{v_H} M_D U_R = \frac{1}{v_H}~ \left( \begin{array}{ccc}
a & \kappa_1 d & \kappa_2 c \\
\kappa_2 d & c & \kappa_1 a \\
\kappa_1 c & \kappa_2 a & d 
\end{array} \right) 
\left( \begin{array}{ccc}
\sqrt{\frac{2}{3}} \cos \theta_R & \frac{1}{\sqrt{3}} & i~\sqrt{\frac{2}{3}} \sin \theta_R \\
- \frac{1}{\sqrt{6}} \cos \theta_R + \frac{1}{\sqrt{2}} \sin \theta_R & \frac{1}{\sqrt{3}} & -i~(\frac{1}{\sqrt{2}} \cos \theta_R + \frac{1}{\sqrt{6}} \sin \theta_R) \\
- \frac{1}{\sqrt{6}} \cos \theta_R - \frac{1}{\sqrt{2}} \sin \theta_R & \frac{1}{\sqrt{3}} & i~(\frac{1}{\sqrt{2}} \cos \theta_R - \frac{1}{\sqrt{6}} \sin \theta_R)
\end{array} \right)  ~~.
\end{eqnarray}

The CP-asymmetry generated by the {\em i-th} heavy Majorana neutrino decaying into a lepton flavor $\alpha$ is: 
\begin{eqnarray}
\epsilon_{N_i} & = & \sum_{\alpha}{ \frac{\Gamma \left(N_i \rightarrow \bar{L}_{\alpha} + H \right)-\Gamma \left(N_i \rightarrow L_{\alpha} + H^* \right)}
{\Gamma \left(N_i \rightarrow \bar{L}_{\alpha} + H \right)+\Gamma \left(N_i \rightarrow L_{\alpha} + H^* \right)}}
\end{eqnarray}
The flavored CP asymmetry $\epsilon_{i}^{\alpha}$ is given by ~\cite{bhupal}:
\begin{eqnarray}
\epsilon_i^{\alpha} & = & \sum_{j \neq i} \frac{Im \left[ (Y_{\nu})_{i \alpha} (Y_{\nu})^*_{j \alpha} \left( Y_{\nu} Y_{\nu}^{\dagger}\right)_{ij} \right] + |\frac{M_i}{M_j}|~ 
Im \left[ (Y_{\nu})_{i \alpha} (Y_{\nu})^*_{j \alpha} \left( Y_{\nu} Y_{\nu}^{\dagger}\right)_{ji} \right]}
{\left( Y_{\nu} Y_{\nu}^{\dagger}\right)_{ii} \left( Y_{\nu} Y_{\nu}^{\dagger}\right)_{jj}} \left(f^{mix}_{ij} + f^{osc}_{ij} \right) 
\end{eqnarray}
where the regulators are: 
\begin{eqnarray}
f^{mix}_{ij} & = & \frac{\left(M_i^2 - M_j^2 \right) |M_i| \Gamma_j}{\left(M_i^2 - M_j^2 \right)^2+  M_i^2 \Gamma_j^2} \nonumber \\
f^{osc}_{ij} & = & \frac{\left(M_i^2 - M_j^2 \right) |M_i| \Gamma_j}{\left(M_i^2 - M_j^2 \right)^2 + \left(|M_i| \Gamma_i + |M_j| \Gamma_j \right)^2 
\frac{det \left[ Re \left(Y_{\nu} Y_{\nu}^{\dagger} \right)\right]}{\left( Y_{\nu} Y_{\nu}^{\dagger}\right)_{ii} \left( Y_{\nu} Y_{\nu}^{\dagger}\right)_{jj}}} ~~~~.
\end{eqnarray}
Here $\Gamma_i$ is the decay width of the {\em i-th} right-handed Majorana neutrino, given at tree level by:  
\begin{eqnarray}
\Gamma_i & = & \frac{|M_i|}{8 \pi}\left( Y_{\nu} Y_{\nu}^{\dagger}\right)_{ii} ~~~~~~~~.
\end{eqnarray}

Taking into account the explicit form of the Yukawa coupling $Y_{\nu}$ and the heavy neutrino masses, the different element of $\Gamma_i$ can be written as: 
\begin{eqnarray}
\Gamma_1 & = & \frac{f \sqrt{1-r+r^2}}{8 \pi v_H^2} \left(|a|^2 + |c|^2 + |\kappa_1|^2 (|c|^2 + |d|^2) + 2 |c|^2 Re (\kappa_1) \right) \nonumber \\
\Gamma_2 & = & \frac{f (1+r) }{8 \pi v_H^2} \left(|c|^2 + |d|^2 + |\kappa_1|^2 (|a|^2 + |d|^2) + 2 |d|^2 Re (\kappa_1) \right) \nonumber \\
\Gamma_3 & = & \frac{f \sqrt{1-r+r^2}}{8 \pi v_H^2} \left(|a|^2 + |d|^2 + |\kappa_1|^2 (|a|^2 + |c|^2) + 2 |a|^2 Re (\kappa_1) \right) \nonumber ~~~~~.\\
\end{eqnarray}

For each $N_i$, one introduces the flavored decay parameter $K_i^{\alpha}$ defined as: 
\begin{eqnarray}
K_i^{\alpha} & = & \frac{\Gamma \left(N_i \rightarrow \bar{L}_{\alpha} + H \right)+\Gamma \left(N_i \rightarrow L_{\alpha} + H^* \right)}{H_N (T= |M_i|)} = 
\frac{\tilde{m}_i^{\alpha}}{m_*} 
\end{eqnarray}
where $H_N (T=|M_i|)$ is the Hubble parameter at temperature $T=|M_i|$ given by: 
\begin{eqnarray}
H_N (T=|M_i|) & = & \sqrt{\frac{4 \pi^3 g_*}{45}} \frac{M_i^2}{M_{Planck}}
\end{eqnarray}
with $g_* = 106.75$ is the number of relativistic degrees of freedom of the Standard Model at high temperatures and $M_{Planck} = 1.2 \times 10^{19}~GeV$ is the planck mass.\\
Here, the effective flavored neutrino mass $\tilde{m}_i^{\alpha}$ is defined as:
\begin{eqnarray}
\tilde{m}_i^{\alpha} & = & v_H^2 \frac{|(Y_{\nu})_{i \alpha}|^2}{|M_i|}
\end{eqnarray}
and $m_*$ is the so-called equilibrium neutrino mass given by: 
\begin{eqnarray}
m_* & = & \frac{16 \pi^{5/2} \sqrt{g_*}}{3 \sqrt{5}} \frac{v_H^2}{M_{Planck}} \sim 1.08 \times 10^{-3} ~eV  ~~~~~~~.
\end{eqnarray}
The washout factor $K_i$, which is a key ingredients for the thermodynamical description of the decays of heavy particles in the early stages of the Universe, is obtained by summing over all flavor, 
\begin{eqnarray}
K_i & = & \frac{\Gamma_i}{H_N (T=|M_i|)} = \sum_{\alpha} K_i^{\alpha} = \frac{\tilde{m}_i}{m_*} 
\end{eqnarray}
where $\tilde{m}_i = v_H^2 \frac{(Y_{\nu} Y_{\nu}^{\dagger})_{ii}}{|M_i|}$ is the effective neutrino mass which can be recast as: 
\begin{eqnarray}
\tilde{m}_1 & = & \frac{1}{f \sqrt{1-r+r^2}} ~\left(|a|^2 + |c|^2 + |\kappa_1|^2 (|c|^2 + |d|^2) + 2 |c|^2 Re (\kappa_1) \right)  \nonumber \\
\tilde{m}_2 & = & \frac{1}{f (1+r)} ~\left(|c|^2 + |d|^2 + |\kappa_1|^2 (|a|^2 + |d|^2) + 2 |d|^2 Re (\kappa_1) \right)  \nonumber \\
\tilde{m}_3 & = & \frac{1}{f \sqrt{1-r+r^2}} ~\left(|a|^2 + |d|^2 + |\kappa_1|^2 (|a|^2 + |c|^2) + 2 |c|^2 Re (\kappa_1) \right) ~~~~~~.
\end{eqnarray}
The effective neutrino mass $\tilde{m}_i$ is a measure of the departure from equilibrium and if $\tilde{m}_i << m_* (\tilde{m}_i >> m_*)$, the asymmetry is weakly (strongly) washed out by the inverse decays. 

The lepton asymmetry is converted by sphaleron effects into a baryon asymmetry. In the strong washout regime, the resulting baryon asymmetry of the Universe is ~\cite{deppisch}: 
\begin{eqnarray}
\eta_B & = & \frac{n_B - n_{\bar{B}}}{n_{\gamma}} = - \frac{28}{51} \frac{1}{27} \frac{3}{2} \sum_{\alpha,i} \frac{\epsilon_i^{\alpha}}{K^{\alpha}_{eff} min \left(z_c, z_{\alpha} \right)}
\end{eqnarray}
where $K^{\alpha}_{eff} = \kappa^{\alpha} \sum_i K_i B_i^{\alpha}$ is the efficiency washout factor with the branching ratios $B_i^{\alpha}$ of each $N_i$ to each 
specific flavor $\alpha$, 
\begin{eqnarray}
B_i^{\alpha} = \frac{|\left(Y_{\nu} \right)_{i \alpha}|^2}{\left( Y_{\nu} Y_{\nu}^{\dagger}\right)_{ii}} ~~~~,
\end{eqnarray} 
$z_c = M_1/T_c$, $T_c \sim 149~GeV$ is the critical temperature below which the sphaleron transitions freeze-out and $z_{\alpha} = 1.25 \ln (25 K^{\alpha}_{eff})$.\\

The factor $\kappa^{\alpha}$, which includes the effect of the real intermediate state subtracted collision terms, is given by: 
\begin{eqnarray}
\kappa^{\alpha} & = & 2 \sum_{i,j (j \neq i)} \frac{Re \left[ (Y_{\nu})_{i \alpha} (Y_{\nu})^*_{j \alpha} \left( Y_{\nu} Y_{\nu}^{\dagger}\right)_{ij} \right] + 
Im \left[ \left( (Y_{\nu})_{i \alpha} (Y_{\nu})^*_{j \alpha} \right)^2  \right]}{Re \left[ \left( Y_{\nu}^{\dagger} Y_{\nu}\right)_{\alpha \alpha} \left\{ \left( Y_{\nu} Y_{\nu}^{\dagger}\right)_{ii} + \left( Y_{\nu} Y_{\nu}^{\dagger}\right)_{jj }\right\} \right]} \left(1 - 2 i \frac{|M_i| - |M_j|}{\Gamma_i + \Gamma_j} \right)^{-1} ~~~~~.
\end{eqnarray}

In the next section, we present a detailed analysis of our work by dividing it into several subsections, first fit of the model's parameters from low energy neutrino data, then compute of the baryon number asymmetry and then find the allowed space for the effective Majorana neutrino mass $m_{ee}$. 
\section{Analysis}
\label{section5}
\subsection{Fit of $A_4$ Model Using Low Energy Neutrino Data}
In order to see more clearly the implications of the scaling invariant Dirac mass matrix and to check the phenomenological predictions of the model, we have performed a numerical analysis. Before going to into detail of the numerical analysis, we would like to stress that we have succeeded in the previous sections to write explicitly  all the Dirac neutrino mass matrix elements $\kappa_1, a, c$ and $d$ in terms of the low energy neutrino oscillation observables. One has just to find the constrained parameter space of the model parameters using the $3 \sigma$ global fit on neutrino data. 
In what follows, we use the $3 \sigma$ range of neutrino mixing angles and mass-squared differences to constrain the parameters of the model and find the correlations 
among them. \\

The present values of these observables, extracted from global analysis of all neutrino oscillation data are shown in Table  \ref{tab:Table 2} \cite{esteban}.
\begin{table}[!h]  
\begin{center}
\begin{tabular}{|c|c|c|} 
\hline
\hline  
Parameter & best fit $\pm 1 \sigma$   & $3 \sigma$ \\ 
\hline
\hline
                   & NH~~~~~IH & NH~~~~~~~IH \\
\hline
$\Delta m_{\odot}^2 [ 10^{-5} eV^2 ]$ & $7.39^{+0.21}_{-0.20}~~~7.39^{+0.21}_{-0.20}$ &  $6.79-8.01~~~6.79-8.01$ \\ 
\hline
$|\Delta m_{atm}^2| [ 10^{-3} eV^2 ]$ & $+2.525^{+0.033}_{-0.032}~~~2.512^{+0.034}_{-0.032}$ & $+2.427-2.625~~~2.412-2.611$ \\       
\hline
$\sin^2 \theta_{12}$               & $0.320^{+0.013}_{-0.012}~~~0.320^{+0.013}_{-0.012}$ &  $0.275-0.350~~~0.275-0.350$ \\ 
\hline
$\sin^2 \theta_{23}$               & $0.580^{+0.017}_{-0.021}~~~0.580^{+0.016}_{-0.020}$    &    $0.411
8-0.627~~~0.423-0.629$    \\
\hline 
$\sin^2 \theta_{13}$               & $0.02241^{+0.00065}_{-0.00065}~~~0.02264^{+0.00066}_{-0.00066}$ &  $0.02045-0.02439~~~0.02068-0.02463$ \\              
\hline
$\delta^0$                           & $215^{+40}_{-29}~~~284^{+27}_{-29}$     &   $125-392~~~196-360$             \\                                  
\hline 
\end{tabular}
\caption{\label{tab:Table 2} Global oscillation analysis with best fit for $\Delta m^2_{\odot},\Delta m^2_{atm},\sin^2 \theta_{12},\sin^2 \theta_{23},
\sin^2 \theta_{13}$ and the $\delta$ upper and/or lower corresponds to normal and/or inverted neutrino mass hierarchy.}
\end{center}
\end{table}  
It is worth noticing that although the neutrino mixing angles and mass-squared differences are known with very good precision, the sign of $\Delta m^2_{atm}$ is unknown. On the other hand, there are two possible orderings allowed by experiments,
\begin{itemize}
\item{ NH with $m_1 < m_2 = \sqrt{m_1^2 + \Delta m^2_{21}} < m_3 = \sqrt{m_1^2 + \Delta m^2_{31}}$ .} \\
\item{ IH with $m_3 < m_1 = \sqrt{m_3^2 + \Delta m^2_{23}-\Delta m^2_{21}} < m_2 = \sqrt{m_3^2 + \Delta m^2_{23}}$ .} \\
\end{itemize}
Moreover the experimental sensitivity to the values of $\delta, \alpha$ and $\beta$ are still limited and the absolute mass scale of neutrinos is unknown. \\
In our analysis, we take as inputs $m_{\small{lightest}}, \Delta m_{\odot}^2 , \Delta m_{atm}^2 $,
$\sin^2 \theta_{12}, \sin^2 \theta_{13}, \sin^2 \theta_{23}, \delta, \alpha$ and  $\beta$. We have nine observables and for each set of input observables, we have generated one million of random sets within $3 \sigma$ intervals, for both normal and inverted hierarchical neutrino masses 
( $m_{lightest} \in [\, 10^{-6}, 0.1 ]\, eV$), mixings, Dirac phase and Majorrana phases between $0$ and $2 \pi$. In addition to the bounds from neutrino oscillation experiments and undetermined Majorana CP phases, the parameters of the model can also get constrained in order to satisfy the cosmological upper limit on the sum of absolute neutrino mases $\sum_{\alpha=e,\mu,\tau} |m_{\nu_{\alpha}}| < 0.17~eV$. \\
The heavy right-handed Majorana mass matrix, which has two parameters $f$ and $r$, is taken to be a real symmetric mass matrix and hence contains two real elements. We set $f = 5~TeV$ and randomly vary the parameter $r$. 
The small variation of the $r$ helps to generate a small mass splitting between the two heavy neutrino of the order of their average decay width and to achieve a successful leptogenesis. 
It turns out that in order to ensure a viable resonant regime, we have randomly to vary $r$ from $10^{-10}$ to $10^{-8}$. Such a choice will reproduce the observed baryon asymmetry compatible with neutrino oscillation. 
By doing so, we have numerically determined the elements of the neutrino Dirac mass matrix $M_D$ and the heavy right-handed neutrino mass matrix $M_R$ consistent with the experimental observations. \\
All points in figures satisfy the neutrino oscillations data in the $3 \sigma$ range. Some of the interesting correlation plots are shown in figure 1. There is a strong correlation between different neutrino parameters of the model with each other for both normal and inverted hierarchies. One can see that the magnitude of the scaling factor $\kappa_1$ is inversely correlated to the magnitude of the parameter $a$. Figure 1 indicates that the parameters of the model have preferences for particular ranges. It is remarkable to find that the viable points for the magnitude of $a, c$ and $d$ are within the $MeV$ or below, whereas most of the points for the scaling factor lie below $10$.
We also find that the low energy neutrino data do not favor $ \kappa_1 = -1$ (i.e. $\kappa_2=0$) which corresponds to a vanishing lightest neutrino mass and that explain the white band shown in the top two plots of figure 1. We can clearly see from the below two plots of figure 1 that neutrino data experiments favor 
$|d| \simeq |c|$ which indicates that the triplet scalars $\phi_{\nu}$ has the vacuum expectation values (vevs) $|v_{\nu 3}| \simeq |v_{\nu 2}| \simeq |v_{\nu 1}|$ which justify our choice of a small perturbations with respect to the exact vacuum alignment. \\
In Figure 2, we have plotted the scaling factor $\kappa_1$ as a function of the parameter $r$. As shown in the first two plots of figure 2, the parameter $|\kappa_1|$ is essentially unconstrained and the whole range $[\,10^{-10}, 10^{-8} ]\,$ of $r$ is allowed. This is consistent with the fact that the magnitude of the parameters of the model are insensitive to the parameter $r$ which is needed for successful leptogenesis. Moreover, we see clearly from the other two plots of figure 2 that the 
$3 \sigma$ range of the neutrino oscillation restricts the phase $\kappa_1$ into three narrow disconnected bands. The allowed values of the phase of $\kappa_1$ are through the ranges $\left[- \pi, -0.65 \pi \right], \left[-0.6 \pi, -0.4 \pi \right]$ and $\left[ 0.55 \pi, \pi \right]$ for both normal and inverted hierarchies. \\
This feature becomes clear from the top two plots of figure 3 where the magnitude of the scaling factor $\kappa_1$ is shown as a function of its phase $\varphi_{\kappa_1}$. The rest plots in figure 3 show the relevant parameter space of the magnitudes as a function of their phases. 
The whole range $[\,-\pi, \pi ]\,$ is allowed. \\  
In Figures 4, the parameters of the model are plotted as a function of the lightest neutrino mass $m_1$ for NH and $m_3$ for IH. We can see clearly that the white band around $|\kappa_1| \simeq 1$ for the whole range attempted (from $10^{-6}~eV$ to $0.1~eV$). While the magnitude of the scaling factor tends to have values less than $3$ for larger values of lightest neutrino mass, the parameters $|a|, |c|$ and $|d|$ can take values in $MeV$ or below in the whole range of $m_{lightest}$. \\ 
For completeness, we have carried out a through numerical analysis study of the parameters of the model as a function of the CP-Majorana phases $\alpha$ and $\beta$, and the Dirac CP-violating phase $\delta$. where the phases are allowed to vary within the full range $[\, 0, 2 \pi ]\,$. We find that the parameters of the model have values over the entire range of CP-phases for both normal and inverted hierarchies. \\

\begin{figure}[hbtp]
\centering
  \begin{tabular}{@{}cc@{}}
  \includegraphics[width=.4\textwidth]{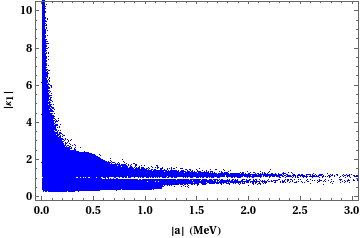} & \includegraphics[width=.4\textwidth]{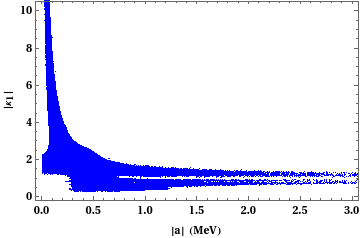} \\
    \includegraphics[width=.4\textwidth]{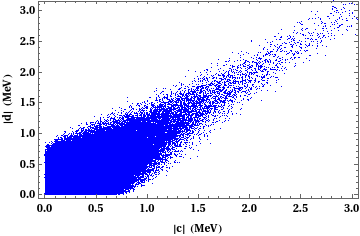} & \includegraphics[width=.4\textwidth]{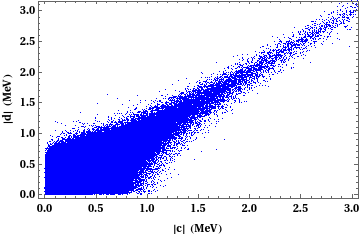} \\
\end{tabular}
\caption{ Correlation between the scaling parameter $|\kappa_1|$ and $|a|$, and correlation between the parameters $|d|$ versus $|c|$ for NH (left panel) and IH (right panel).}
\end{figure}

\begin{figure}[hbtp]
\centering
  \begin{tabular}{@{}cc@{}}
  \includegraphics[width=.4\textwidth]{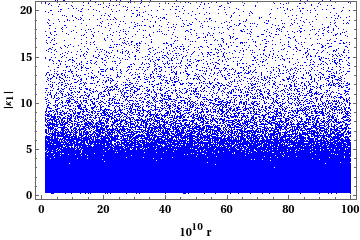} & \includegraphics[width=.4\textwidth]{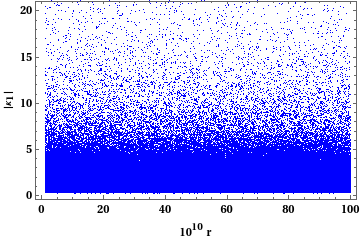} \\
 \includegraphics[width=.4\textwidth]{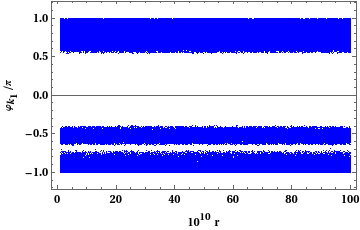} & \includegraphics[width=.4\textwidth]{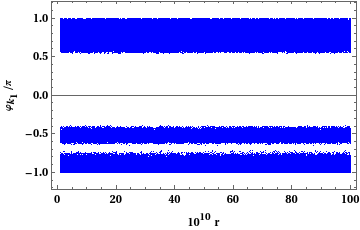}  \\
\end{tabular}
\caption{Plot of the scale factor $\kappa_1$ versus the parameter $r$ for NH (left panel) and IH (right panel).}
\end{figure}

\begin{figure}[hbtp]
\centering
  \begin{tabular}{@{}cc@{}}
  \includegraphics[width=.4\textwidth]{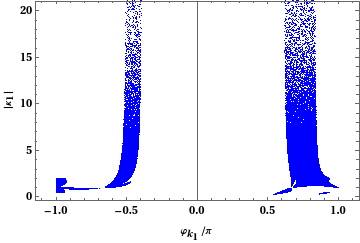} & \includegraphics[width=.4\textwidth]{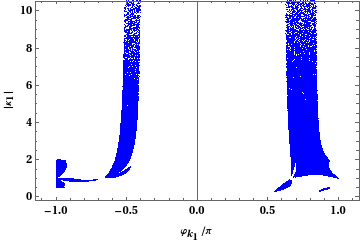} \\
  \includegraphics[width=.4\textwidth]{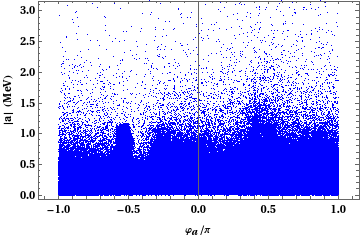} &  \includegraphics[width=.4\textwidth]{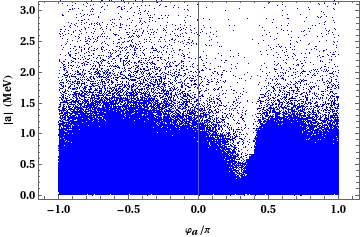}\\                                          
    \includegraphics[width=.4\textwidth]{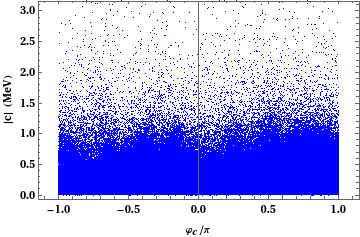} & \includegraphics[width=.4\textwidth]{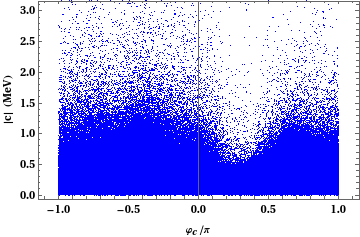} \\
    \includegraphics[width=.4\textwidth]{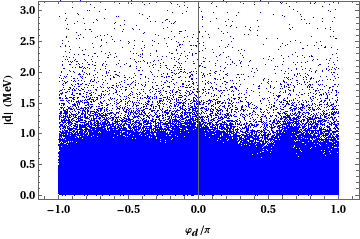} & \includegraphics[width=.4\textwidth]{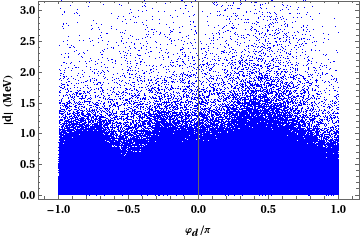}\\
\end{tabular}
\caption{Plot of the parameters of the model $|\kappa_1, |a|, |c|$ and $|d|$ versus their phases for NH (left panel) and IH (right panel).}
\end{figure}

\begin{figure}[hbtp]
\centering
  \begin{tabular}{@{}cc@{}}
  \includegraphics[width=.4\textwidth]{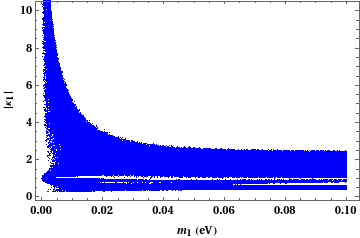} & \includegraphics[width=.4\textwidth]{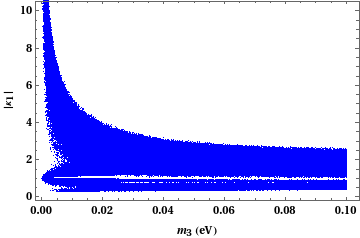} \\
    \includegraphics[width=.4\textwidth]{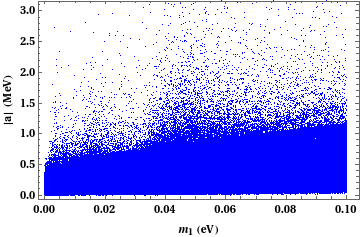} & \includegraphics[width=.4\textwidth]{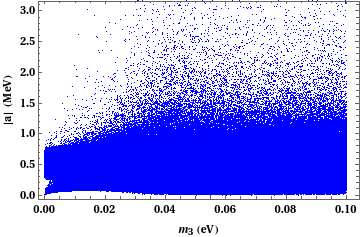} \\
    \includegraphics[width=.4\textwidth]{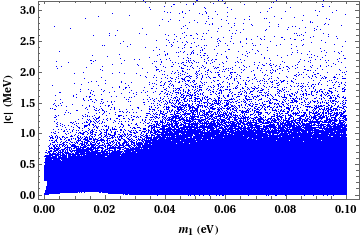} & \includegraphics[width=.4\textwidth]{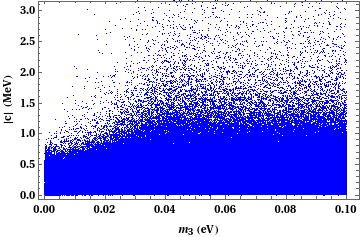} \\
    \includegraphics[width=.4\textwidth]{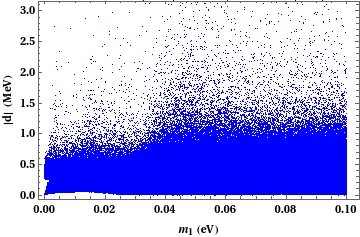} & \includegraphics[width=.4\textwidth]{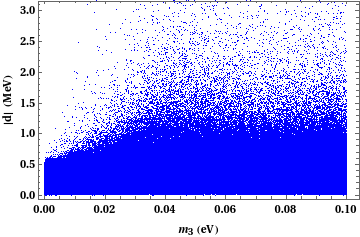} \\
\end{tabular}
\caption{Plot of the parameters of the model $|\kappa_1, |a|, |c|$ and $|d|$ versus the lightest neutrino mass for NH (left panel) and IH (right panel).}
\end{figure}
\clearpage
\subsection{Numerical Results for Baryogenesis via Resonant Leptogenesis} 
The out of equilibrium decays of the heavy right handed Majorana neutrinos generate a lepton asymmetry, which is subsequently partially converted into a baryon asymmetry through by spheleron processes at high temperatures.The number densities of baryons anti-baryons compared to that of photons in the universe is tiny. The observed baryon number asymmetry today is: 
\begin{eqnarray}
\eta_B & = & \frac{n_B - n_{\bar{B}}}{n_{\gamma}} = \left(6.04 \pm 0.08 \right) \times 10^{-10}
\end{eqnarray}
where $n_{B (\bar{B})}$ is the number of baryons (anti-baryons) density, $n_{\gamma}= 2 T^3 \zeta(3)/\pi^2$ is the number density of photons and $\zeta(x)$ is the Riemann zeta function with $\zeta(3) = 1.20206$. \\

Before going into details of the numerical analysis, we would like to stress that we have succeeded in the previous sections to write explicitly all the Dirac neutrino mass matrix elements $\kappa_1, a, c$ and $d$ in terms of the low energy neutrino oscillation observables. In addition, we have found the allowed ranges of the model's parameters using the $3 \sigma$ global fit on neutrino oscillation data. \\

We now turn to the calculation related to baryogenesis via resonant leptogenesis. We consider a scenario with three nearly degenerate heavy right-handed Majorana neutrinos, for which the masses are around $5~TeV$. Computation of the baryon number asymmetry requires full information on the parameters of the Dirac and Majorana mass matrices. After finding the parameter space of the parameters $\left(\kappa_1, a, c, d \right)$ along with the ranges of $f$ and $r$ consistent with the $3 \sigma$ global fit, we feed the one million data sets to calculate the baryogenesis via resonant leptogenesis for each point of the data sets. Our aim is to delineate more the viable parameter space of the model's parameters by considering both case normal and inverted hierarchies. As described in section 4, we have expressed the neutrino Yukawa coupling $Y_{\nu}$ in terms of the model's parameters $\kappa_1, a, c, d$ and $r$ which are consistent with the low energy neutrino data. \\

To start with, we have plotted in figure 5 the numerical estimates of the baryon number asymmetry $\eta_B$ as a function of the magnitude of model's parameters. The horizontal solid lines denote the phenomenologically allowed region of the observed baryon asymmetry. One can see that we have obtained further restriction on the parameter space of $\kappa_1, a, c$ and $d$ for both NH (right panel) and IH (left panel) hierarchies. Interestingly, we observe that successful baryogenesis is realized within preferred regions of the model's parameters. In particular, there are many points in the parameter space of $\kappa_1, a, c$ and $d$ that predict 
the observed baryon number asymmetry. \\
Figure 6 shows the variation of the baryon number asymmetry over the full range $\left[ -\pi, \pi \right]$ of the argument $\varphi_{\kappa_1}, \varphi_a, \varphi_c$ and $\varphi_d$ of the model's parameters. Extreme values for the argument of the scaling factor $\varphi_{\kappa_1}$ are consistent with the observed baryon asymmetry for both NH and IH hierarchies. On the other hand, other plots in figure 9 show preferred ranges for $\varphi_a, \varphi_c$ and $\varphi_d$. These ranges change if one consider NH (right panel) or IH (left panel) hierarchies. \\  

The prediction of $\eta_B$ in resonant leptogenesis as a function of the CP violating Majorana phases $\alpha$ and $\beta$, Dirac CP phase $\delta$ and lightest neutrino mass $m_1$ (NH) and $m_3$ (IH) are shown in figure 7. As can be seen in figure 10, for successful resonant leptogenesis, the Majorana CP phases are required to take values and ranges. However, as can be seen in the figure 10, all values of the Dirac CP phase reproduce the correct value of the baryon asymmetry and there is no preference. On the other hand, the last two plots in figure 10 show the variation of $\eta_B$ as a function of the lightest neutrino mass. As one can see from the plots, values of $m_1$ (NH) and $m_3$ (IH) above $0.005~eV$ are preferred. \\

We can see that we have appreciable resonant leptogenesis compatible with baryon asymmetry for the space parameter of $\kappa_1, a, c$ and $d$ consistent with low energy neutrino oscillation data.\\

\begin{figure}[hbtp]
\centering
  \begin{tabular}{@{}cc@{}}
  \includegraphics[width=.4\textwidth]{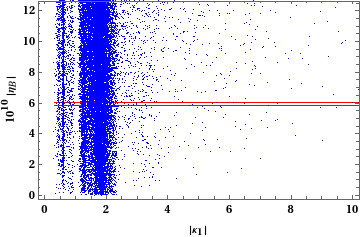} & \includegraphics[width=.4\textwidth]{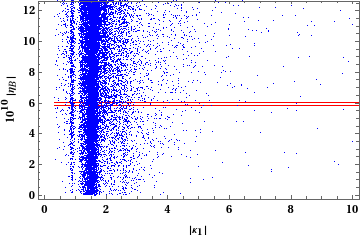} \\
    \includegraphics[width=.4\textwidth]{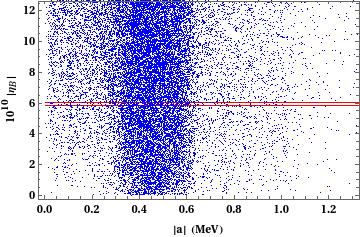} & \includegraphics[width=.4\textwidth]{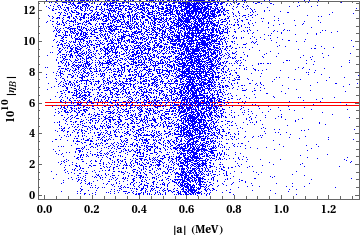} \\
    \includegraphics[width=.4\textwidth]{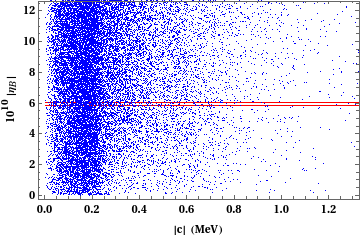} & \includegraphics[width=.4\textwidth]{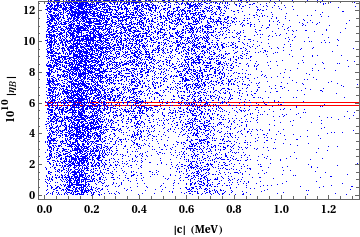} \\
    \includegraphics[width=.4\textwidth]{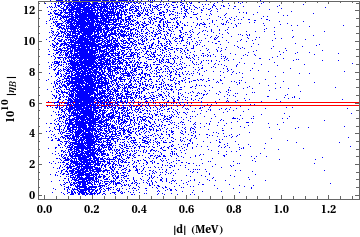} & \includegraphics[width=.4\textwidth]{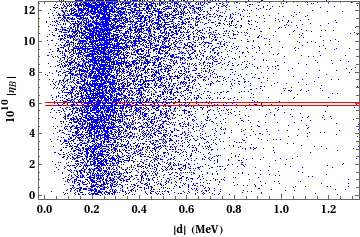} \\
    \includegraphics[width=.4\textwidth]{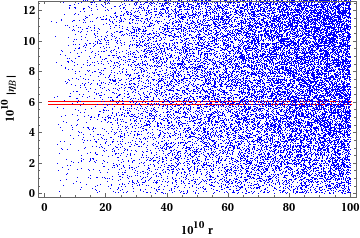} & \includegraphics[width=.4\textwidth]{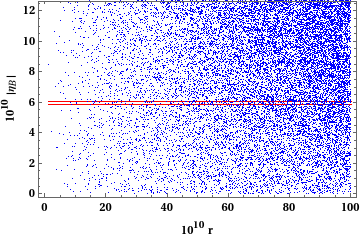} \\
\end{tabular}
\caption{Plot of the CP asymmetry $\eta_B$ versus the parameters of the model $|\kappa_1|, |a|, |c|, |d|$  and $r$ for NH (left panel) and IH (right panel).}
\end{figure}

\begin{figure}[hbtp]
\centering
  \begin{tabular}{@{}cc@{}}
  \includegraphics[width=.4\textwidth]{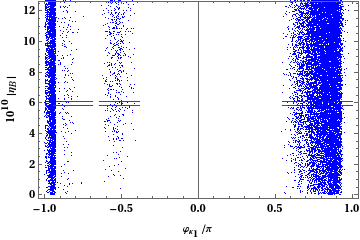} & \includegraphics[width=.4\textwidth]{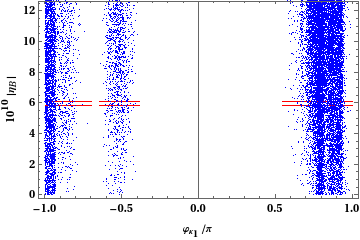} \\
    \includegraphics[width=.4\textwidth]{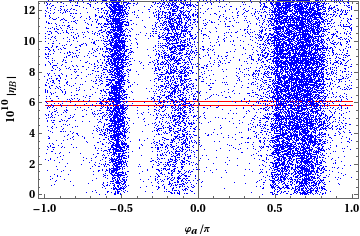} & \includegraphics[width=.4\textwidth]{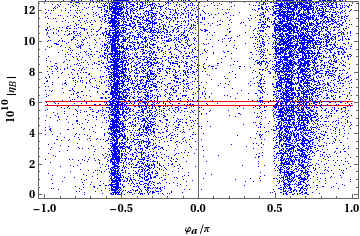} \\
    \includegraphics[width=.4\textwidth]{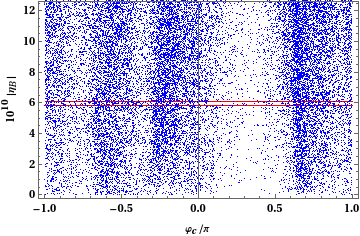} & \includegraphics[width=.4\textwidth]{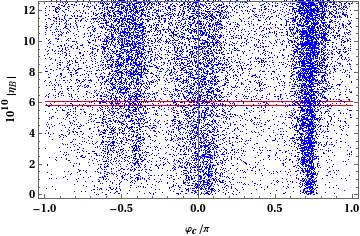} \\
    \includegraphics[width=.4\textwidth]{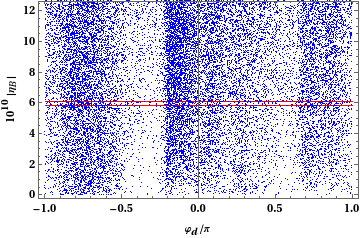} & \includegraphics[width=.4\textwidth]{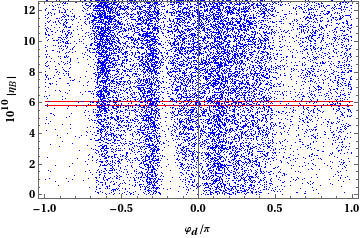} \\
\end{tabular}
\caption{Plot of the CP asymmetry $\eta_B$ versus the phases of the model for NH (left panel) and IH (right panel).}
\end{figure}

\begin{figure}[hbtp]
\centering
  \begin{tabular}{@{}cc@{}}
  \includegraphics[width=.4\textwidth]{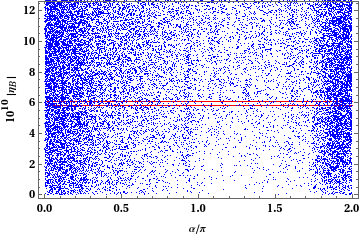} & \includegraphics[width=.4\textwidth]{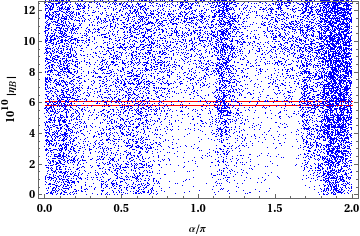} \\
    \includegraphics[width=.4\textwidth]{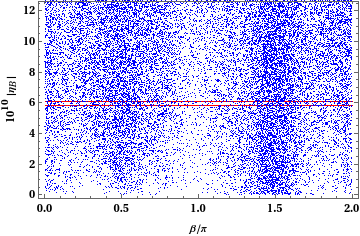} & \includegraphics[width=.4\textwidth]{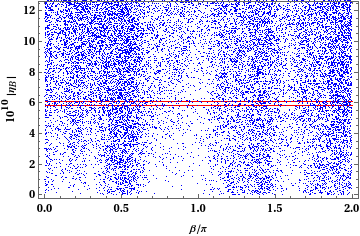} \\
    \includegraphics[width=.4\textwidth]{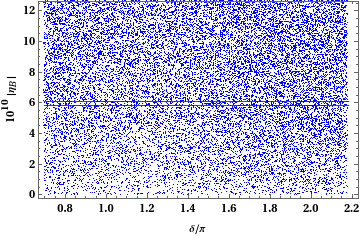} & \includegraphics[width=.4\textwidth]{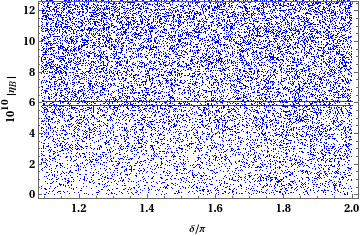} \\
    \includegraphics[width=.4\textwidth]{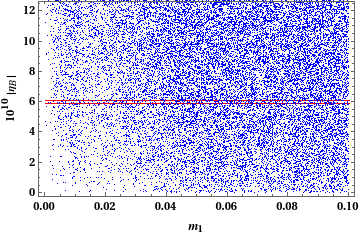} & \includegraphics[width=.4\textwidth]{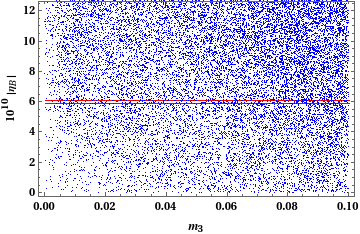} \\
\end{tabular}
\caption{Plot of the CP asymmetry $\eta_B$ versus Majorana phases $\alpha$ and $\beta$, Dirac phase $\delta$ and lightest mass of the model for NH (left panel) and IH (right panel).}
\end{figure}
\clearpage
\subsection{Constraints from Neutrinoless Double Beta Decay}
The effective neutrino mass governing the $0 \nu 2 \beta$ decay is given by: 
\begin{eqnarray}
| m_{ee} | & = & | \sum_{i=1}^3 \mu_i U_{ei}^2| 
\end{eqnarray}
and corresponds to the (11) entry of the neutrino mass matrix $M_{\nu}$: 
\begin{eqnarray}
| m_{ee} | & = & | M_{\nu~11}|  \nonumber\\
& = & \frac{1}{f (1+r^3)} ~| - a^2 + 2 \kappa_1 (\kappa_1 +1) c d + r \left( 2 \kappa_1 a d + (\kappa_1 +1)^2 c^2 \right) 
+ r^2 \left( 2 (\kappa_1 +1 ) a c - \kappa_1^2 d^2 \right)|
\end{eqnarray}
The results for the baryon asymmetry $\eta_B$ as a function of $|m{ee}|$ are depicted in figure 8. The horizontal solid lines correspond to the allowed region of the observed baryon asymmetry. The important feature of the model is the existence of a lower bound $|m_{ee}| > 0.025~eV$ for NH and $|m_{ee}| > 0.10~eV$ for IH which is within the sensitivity reach of future double decay experiments~\cite{gerda}.
\begin{figure}[hbtp]
\centering
  \begin{tabular}{@{}cc@{}}
  \includegraphics[width=.4\textwidth]{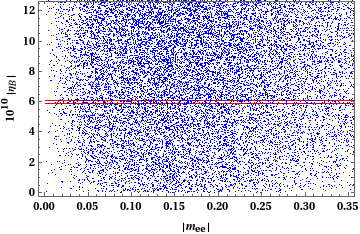} & \includegraphics[width=.4\textwidth]{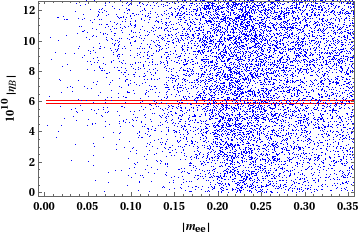} \\
\end{tabular}
\caption{Plot of the CP asymmetry $\eta_B$ versus $m_{ee}$ for NH (left panel) and IH (right panel).}
\end{figure}
\section{Conclusion}
In this paper, we have considered a new scaling structure for the Dirac neutrino mass matrix. In the framework of type-I seesaw, we have reconstructed analytically all the elements of the Dirac mass matrix from the low energy neutrino oscillation data. 
A concrete model, in type-I seesaw mechanism, based on $A_4$ flavor symmetry, has been proposed to realize the new scaling ansatz on the neutrino Dirac mass matrix. 
The model imposes a relation between the two scaling parameters $\kappa_1$ and $\kappa_2$. From equation (19), other relations between the two scaling factors have been found that lead to different phenomenological models which were not considered in the present work.
In this $A_4$ flavor symmetry model, the heavy right-handed Majorana neutrinos are quasi-degenerate at TeV mass scales. We have calculated the Yukawa couplings necessary to generate the lepton asymmetry. Successful leptogenesis has been achieved due to the mass splitting between the TeV scale sterile neutrino providing a resonant enhancement of the asymmetry.
We have constrained the model's parameters by doing a random scan, using the low energy neutrino oscillations data in the $3 \sigma$ range for both normal and inverted hierarchies, to generate the allowed parameter space of the model. It has been found that the parameter space of the neutrino Dirac mass matrix elements lies near or below the MeV region. \\
After finding the model and neutrino parameters consistent with the low energy neutrino oscillations data, we took an interesting endeavor to further constrain the  parameter space of the model using the observed baryon number asymmetry in the Universe. In order to exclude regions of the model parameter space, we fed the one million data sets consistent with neutrino oscillation data to further constrain the model's parameters. By doing so, we have examined the possibility for simultaneous explanation of both the neutrino oscillation data and the observed baryon number asymmetry in the Universe. It has been found that further restrictions on the parameter space of the model are required to explain both the correct neutrino data and the baryon asymmetry vis resonant leptogenesis. moreover, we have shown that the allowed space for the effective Majorana neutrino mass $m_{ee}$ is also constrained in order to account for the observed baryon asymmetry. \\
The scaling ansatz on the neutrino Dirac mass matrix considered here is consistent with the low energy neutrino oscillations data, the observed baryon asymmetry and neutrinoless double beta decay. \\
Other patterns of scaling ansatz on the neutrino Dirac mass matrix could be studied which could be excluded or allowed by the experimental data and might lead to rich phenomenology. We will leave the important aspects of these patterns in future work. \\

\section*{Acknowledgments}
We would like to thank W. Rodejohann and D. Borah for reading the manuscript.

\end{document}